\pdfoutput=1
\documentclass[11pt]{article}

\usepackage[letterpaper,margin=1in]{geometry}
\usepackage[utf8]{inputenc}
\usepackage[T1]{fontenc}
\usepackage{newtxtext,newtxmath}
\usepackage{microtype}
\usepackage{setspace}
\usepackage{authblk}
\usepackage{amsmath,bm}
\usepackage{graphicx}
\usepackage{booktabs}
\usepackage{multirow}
\usepackage{array}
\usepackage{tabularx}
\usepackage{caption}
\usepackage{placeins}
\usepackage[super,sort&compress,numbers]{natbib}
\usepackage[colorlinks=true,linkcolor=black,citecolor=black,urlcolor=blue]{hyperref}

\graphicspath{{./}}
\title{\textbf{Text-guided flow matching enables sample-efficient crystal structure generation}}
\hypersetup{
  pdftitle={Text-guided flow matching enables sample-efficient crystal structure generation},
  pdfauthor={Wentao Li},
  pdfsubject={Text-conditioned flow matching for crystal structure prediction and de novo generation},
  pdfkeywords={crystal generation, flow matching, text conditioning, materials informatics, TFMat}
}

\author[1]{Wentao Li}
\affil[1]{Tsinghua University}

\date{}

\begin{document}

\maketitle

\begin{abstract}
Crystal generators can now propose periodic structures, but their control interfaces remain poorly matched to the mixed descriptors used in materials design. Text provides a compact way to combine composition, symmetry, prototype and property cues, yet it has not been clear whether such information can steer flow-based crystal generation. Here we introduce TFMat, a text-conditioned flow-matching framework that uses structured materials language as a semantic prior for a CrystalFlow generator. Across Perov-5, Carbon-24 and MP-20 crystal structure prediction benchmarks, TFMat improves one-candidate match rates over CrystalFlow and reaches a 92.04\% MP-20 match rate with 20 candidates; in de novo generation, it improves element-count and density distribution alignment while retaining coarse property consistency in composition-selected outputs. These results position structured text as an inspectable control layer for translating human-readable materials intent into candidate crystals for downstream simulation and validation.
\end{abstract}

\section*{Introduction}
Generative modeling has changed the computational search for crystalline materials from enumerating known prototypes to proposing new periodic structures under chemical and functional constraints. The opportunity is large because stable crystals occupy a sparse subset of an enormous space of compositions, lattices and atomic arrangements.\cite{Oganov2019CSPReview} High-throughput materials databases and community benchmark tools now provide the infrastructure for data-driven crystal design,\cite{Jain2013,Curtarolo2012,Kirklin2015,Draxl2019NOMAD,Choudhary2020JARVIS,Ong2013Pymatgen,Ward2016Magpie,Ward2018Matminer,Dunn2020Matbench} while graph neural networks and large-scale discovery pipelines have shown that learned atomic environments and interatomic representations can support accurate structure-property modeling.\cite{Butler2018,XieGrossman2018,Schutt2018SchNet,Chen2019MEGNet,Choudhary2021ALIGNN,Chen2022M3GNet,Deng2023CHGNet,Merchant2023GNoME} The most useful generative model must therefore do more than populate a benchmark distribution: it must respond to the kinds of constraints that define an actual materials question. Such control becomes especially important when candidate structures must be ranked, relaxed and interpreted across downstream computations. The remaining challenge is not simply to generate valid crystals, but to make generation controllable in the language of materials design.

Recent crystal generative models have addressed this challenge mainly by strengthening the geometric model. Conditional graph generators and physics-guided adversarial models established that chemical and symmetry constraints can improve validity,\cite{Gebauer2022CGSchNet,Kim2020CubicGAN,Zhao2023PGCGM} and diffusion or flow-based methods then made it possible to jointly model periodic coordinates, lattices and atom types.\cite{Xie2022CDVAE,Jiao2023DiffCSP,Luo2023SyMat,Luo2024CondCDVAE,Miller2024FlowMM,Luo2025CrystalFlow} Other approaches introduced autoregressive crystal strings, language-model base distributions and large-scale property-conditioned diffusion.\cite{Antunes2024CrystaLLM,Sriram2024FlowLLM,Zeni2025MatterGen} Together these methods have clarified how to respect periodic geometry, but their control variables are often chosen for algorithmic convenience rather than for expressiveness. This gap is methodological as much as architectural, because realistic constraints rarely arrive as isolated labels. In parallel, materials natural-language processing has shown that chemical entities, synthesis recipes and property relations can be extracted from text using parsers, word embeddings and scientific language models.\cite{Swain2016ChemDataExtractor,Kononova2019Recipes,Tshitoyan2019Embeddings,Devlin2019BERT,Beltagy2019SciBERT,Gupta2022MatSciBERT,Shetty2023MaterialsBERT,Song2023MatSciNLP} These advances suggest that crystallographic metadata and language representations could become a shared control layer between human hypotheses, databases and generative models.

This control layer is still underdeveloped. Most crystal generators are unconditional, conditioned only on composition, or driven by a small number of scalar labels. Such inputs are convenient for benchmarks, but they compress materials intent into a narrow representation: a scientist may know a formula, prototype family, space-group tendency, stability range and electronic-property expectation simultaneously, and these descriptors are often mixed across structured database fields and written descriptions. Text-guided diffusion models such as TGDMat and Chemeleon have recently shown that prompt representations can influence periodic materials generation,\cite{Das2025TGDMat,Park2025Chemeleon} and FlowLLM has explored language models as base distributions for flow-based generation.\cite{Sriram2024FlowLLM} However, it remains unclear whether a compact, frozen materials-language prior can improve a continuous crystal flow without replacing its geometric backbone. This is a more stringent question than adding another conditioning channel: if language merely repeats formula labels, its apparent benefit may reflect reconstruction from highly informative metadata rather than useful semantic control. The distinction matters for closed-loop and agentic materials workflows, where autonomous systems must translate research intent into candidates, pass them to simulation or synthesis modules, and keep the decision trail legible to researchers.\cite{MacLeod2020SelfDriving,Szymanski2023ALab,Boiko2023Coscientist} A useful text interface should therefore improve sample-efficient structure recovery now, expose where prompt information is most helpful, and remain honest about the boundary between controlled generation and open-ended discovery.

\begin{figure}[!tbp]
\centering
\includegraphics[width=\textwidth]{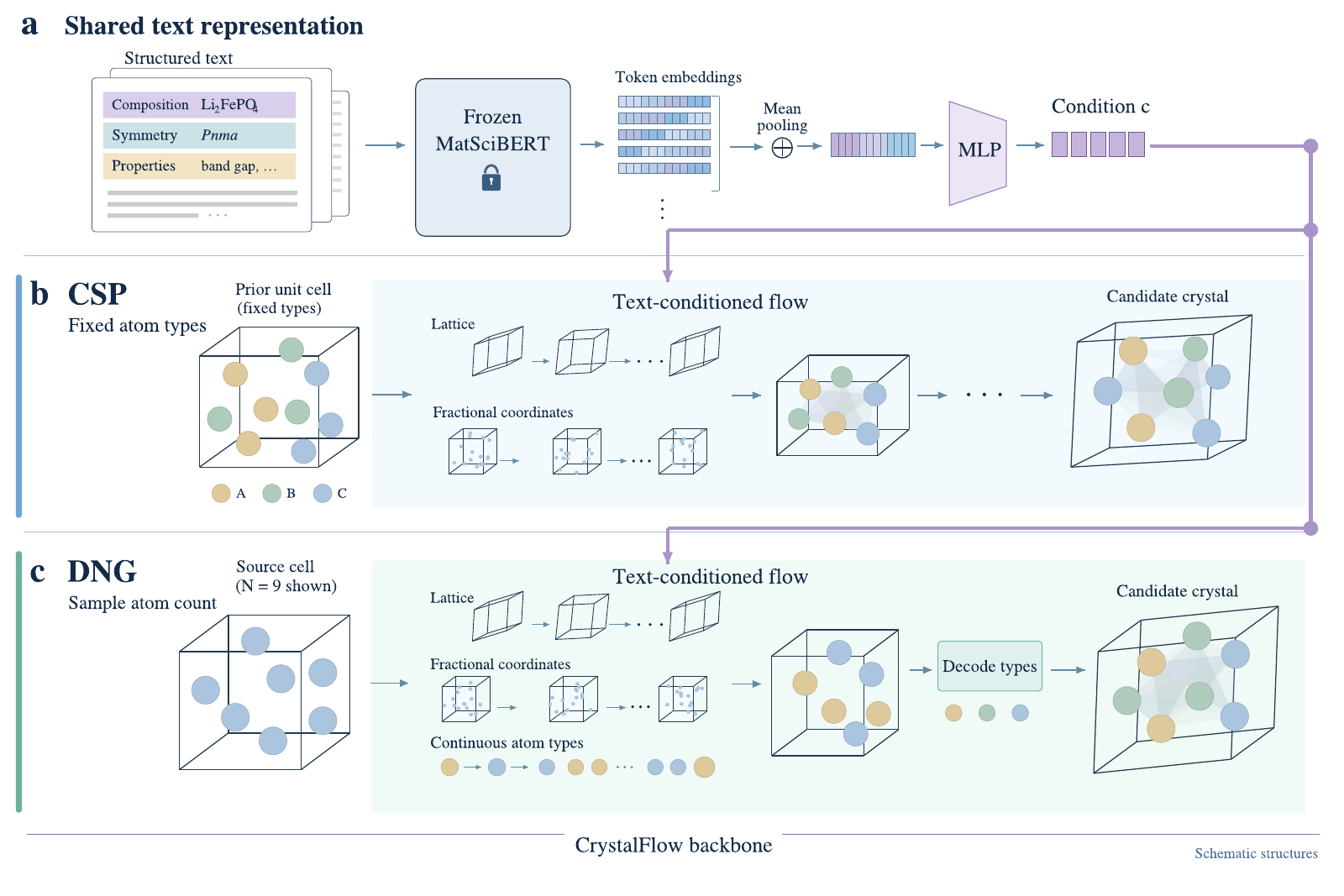}
\caption{\textbf{Overview of the TFMat framework.} \textbf{a}, Structured materials text is encoded by frozen MatSciBERT, mean-pooled and projected to a shared condition vector. \textbf{b}, In crystal structure prediction (CSP), fixed atom types and composition are combined with the text condition while the flow transports lattice parameters and fractional coordinates towards a candidate crystal. \textbf{c}, In de novo generation (DNG), the atom count is sampled and atom types, lattice parameters and fractional coordinates are generated jointly before discrete atom-type decoding. Both regimes retain the CrystalFlow geometric backbone.}
\label{fig:tfmat-overview}
\end{figure}

Here we introduce TFMat, a text-guided extension of CrystalFlow designed to test that premise directly. TFMat keeps the flow-matching crystal generator as the central geometric model and adds a condition vector obtained from frozen MatSciBERT embeddings of structured materials prompts. The prompts are database-derived rather than free-form instructions; they include formula or composition information, symmetry descriptors and selected scalar properties, but not lattice vectors or fractional coordinates. This design isolates the contribution of the text prior more cleanly than replacing the entire generator, while keeping the task close to metadata-rich settings encountered in materials databases. We evaluate TFMat in composition-fixed crystal structure prediction (CSP) and text-conditioned de novo generation (DNG). In CSP, TFMat improves one-sample match rates over CrystalFlow on Perov-5, Carbon-24 and MP-20, with the largest gains on the constrained perovskite and carbon-network benchmarks, and the MP-20 gain persists under a 20-candidate evaluation budget. In DNG, TFMat does not dominate every validity metric, but it improves element-count and density distribution alignment and retains coarse formation-energy and band-gap cues in a composition-selected diagnostic subset. Fig.~\ref{fig:tfmat-overview} summarizes how the shared text representation conditions these two generation regimes. These results support a bounded conclusion: structured language descriptors can bias a flow-based crystal generator towards more relevant regions of crystal space, providing a practical interface for future materials discovery pipelines.

\section*{Results}
\subsection*{Structured prompts define two generation regimes}
We first ask whether text conditioning helps under benchmark conditions that separate structural recovery from open-ended generation. The evaluation uses three standard crystal-generation datasets with increasing chemical and geometric diversity.\cite{Xie2022CDVAE,Jiao2023DiffCSP,Das2025TGDMat} Perov-5 is a constrained five-atom perovskite benchmark and mainly tests recovery of a well-defined prototype family. Carbon-24 contains larger carbon unit cells and tests covalent-network generation in a chemically simple but geometrically demanding setting. MP-20 is a broad Materials Project-derived benchmark of inorganic crystals with up to 20 atoms per unit cell, and is therefore the most diverse dataset considered here.

The same conditioning idea is evaluated in two regimes. In CSP, atom types and composition are fixed while the model predicts lattice parameters and fractional coordinates; this directly tests whether text improves recovery of a known target structure. In DNG, atom types, lattice and fractional coordinates are generated jointly after sampling the number of atoms, so the output is a text-conditioned crystal population rather than a single reconstruction. Throughout, TFMat receives structured prompts derived from database metadata, including formula, symmetry descriptors and selected property values where available. These prompts make the control signal reproducible, but they also define the scope of the claim: TFMat is evaluated as a structured text-conditioned generator, not as an unrestricted natural-language discovery system.

\subsection*{Text conditioning improves one-sample crystal structure prediction}
The clearest test of sample efficiency is the one-sample CSP regime, where each test composition receives only one generated candidate. If the text condition is useful, it should increase the chance that this single trajectory enters the correct structural basin. Table~\ref{tab:evo1} shows that TFMat does so consistently. Relative to CrystalFlow, the match rate increases from 53.69\% to 91.33\% on Perov-5, from 15.02\% to 46.40\% on Carbon-24 and from 67.65\% to 77.80\% on MP-20.

\begin{figure*}[!tbp]
\centering
\includegraphics[width=0.94\textwidth,height=0.48\textheight,keepaspectratio]{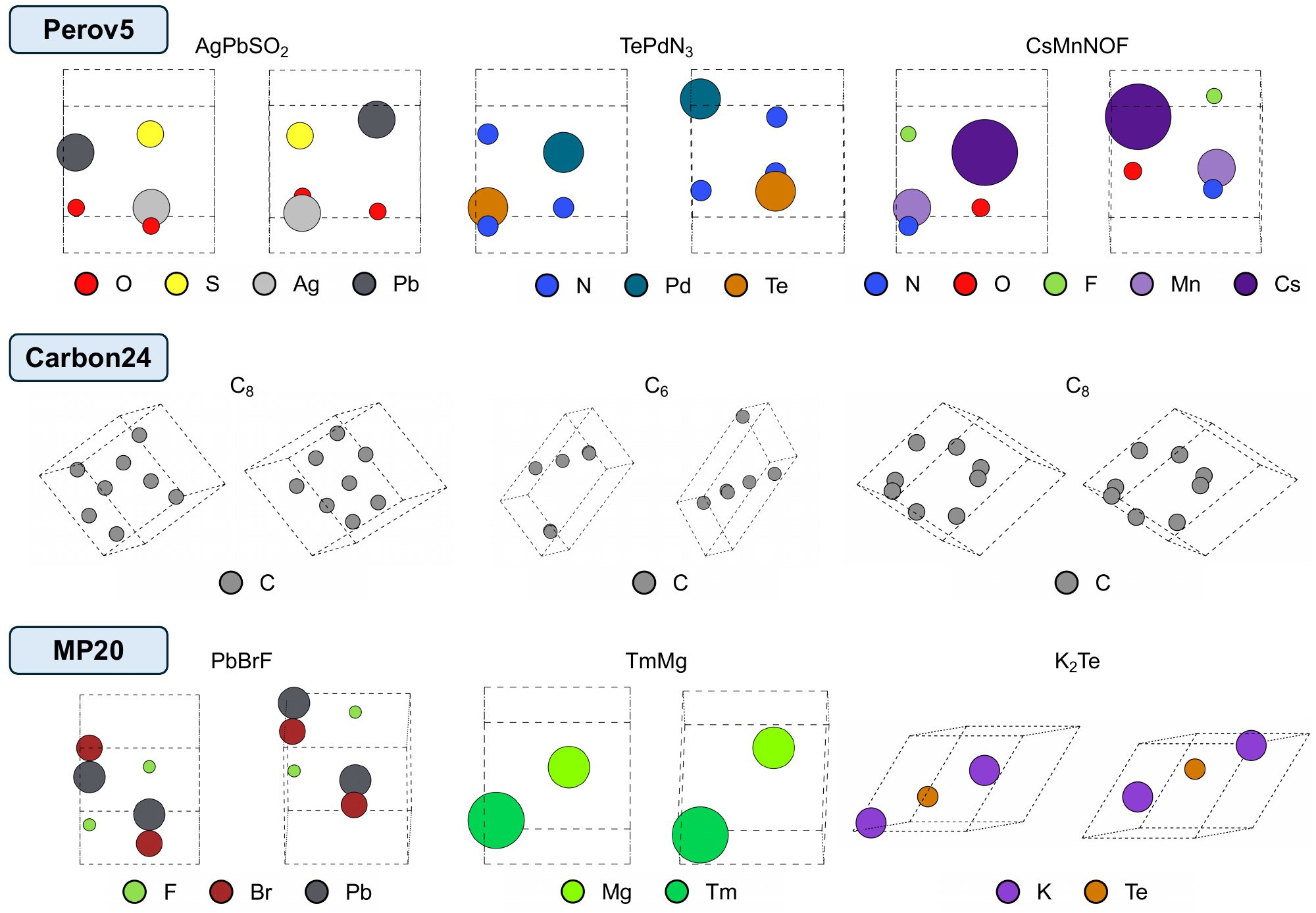}
\caption{\textbf{Representative crystal structure predictions from TFMat.} Each pair shows a ground-truth crystal (left) and a TFMat prediction (right). Atom species are distinguished by color and unit cells are outlined in black.}
\label{fig:csp-qualitative}
\end{figure*}

This improvement is not confined to the easiest benchmark. Among the methods shown in Table~\ref{tab:evo1}, TFMat gives the highest one-sample match rate on all three datasets. The Carbon-24 result is particularly informative because the chemistry is fixed but the covalent networks are geometrically diverse; TFMat raises match rate by 31.38 percentage points and lowers RMSE from 0.3095 to 0.1353. On MP-20, TFMat improves structural retrieval even though its RMSE for matched structures is not the lowest. Representative matched structures are shown in Fig.~\ref{fig:csp-qualitative}. The result therefore points to a specific role for text: it helps select plausible basins of crystal space, while final geometric refinement remains a separate challenge.

\begin{table*}[!htbp]
\centering
\caption{\textbf{Single-sample CSP comparison across Perov-5, Carbon-24 and MP-20.} Best values are bold and second-best values are underlined. Published baseline values are from TGDMat.\cite{Das2025TGDMat} Match rates are percentages; lower RMSE is better.}
\label{tab:evo1}
\small
\setlength{\tabcolsep}{4pt}
\resizebox{\textwidth}{!}{%
\begin{tabular}{lcccccc}
\toprule
Method & Perov-5 Match Rate $\uparrow$ & Perov-5 RMSE $\downarrow$ & Carbon-24 Match Rate $\uparrow$ & Carbon-24 RMSE $\downarrow$ & MP-20 Match Rate $\uparrow$ & MP-20 RMSE $\downarrow$ \\
\midrule
P-cG-SchNet   & 48.22 & 0.4179 & 17.29 & 0.3846 & 15.39 & 0.3762 \\
CDVAE         & 45.31 & 0.1138 & 17.09 & 0.2969 & 33.90 & 0.1045 \\
DiffCSP       & 52.02 & 0.0760 & 17.54 & 0.2759 & 51.49 & 0.0631 \\
TGDMat (Short) & 56.54 & 0.0583 & 24.13 & 0.2424 & 52.22 & \underline{0.0597} \\
TGDMat (Long)  & \underline{90.46} & \textbf{0.0203} & \underline{44.63} & \underline{0.2266} & 55.15 & \textbf{0.0572} \\
CrystalFlow   & 53.69 & 0.0953 & 15.02 & 0.3095 & \underline{67.65} & 0.0791 \\
TFMat (Ours)  & \textbf{91.33} & \underline{0.0449} & \textbf{46.40} & \textbf{0.1353} & \textbf{77.80} & 0.0868 \\
\bottomrule
\end{tabular}%
}
\end{table*}

\subsection*{Repeated sampling confirms the MP-20 retrieval gain}
Repeated sampling tests whether the one-sample gain persists when the model is allowed to propose multiple candidates. In the 20-candidate MP-20 evaluation, a structure is counted as matched if at least one candidate matches the reference. TFMat again gives the highest match rate among the compared methods, reaching 92.04\% compared with 85.38\% for CrystalFlow and 82.02\% for TGDMat (Long) (Table~\ref{tab:mp20-evo20}).

The RMSE ranking defines the boundary of this claim. TGDMat (Short) reports the lowest RMSE at 0.0443, followed by TGDMat (Long) at 0.0483 and DiffCSP at 0.0492. TFMat therefore contributes most strongly to retrieval of the correct basin, rather than to the final geometric refinement of matched candidates. This distinction is practically important: in screening workflows, a candidate in the right basin can be passed to relaxation, whereas relaxation cannot rescue a sample that never reaches a relevant structural family.

\begin{table}[!htbp]
\centering
\caption{\textbf{MP-20 CSP comparison with 20 candidates.} Best values are bold and second-best values are underlined. Published baseline values are from TGDMat.\cite{Das2025TGDMat} Match rates are percentages.}
\label{tab:mp20-evo20}
\small
\setlength{\tabcolsep}{6pt}
\begin{tabular}{lcc}
\toprule
Method & MP-20 Match Rate $\uparrow$ & MP-20 RMSE $\downarrow$ \\
\midrule
P-cG-SchNet    & 32.64 & 0.3018 \\
CDVAE          & 66.95 & 0.1026 \\
DiffCSP        & 77.93 & 0.0492 \\
TGDMat (Short) & 80.97 & \textbf{0.0443} \\
TGDMat (Long)  & 82.02 & \underline{0.0483} \\
CrystalFlow    & \underline{85.38} & 0.0614 \\
TFMat (Ours)   & \textbf{92.04} & 0.0665 \\
\bottomrule
\end{tabular}
\end{table}

\subsection*{Text-conditioned de novo generation reshapes generated populations}
The DNG task is harder because the model must generate composition, lattice and coordinates jointly. Here the question changes from exact recovery to whether structured text can shape the generated population. TFMat's main effect in this setting is distributional rather than a uniform increase in every validity metric.

\begin{figure}[!htbp]
\centering
\includegraphics[width=\textwidth,height=0.62\textheight,keepaspectratio]{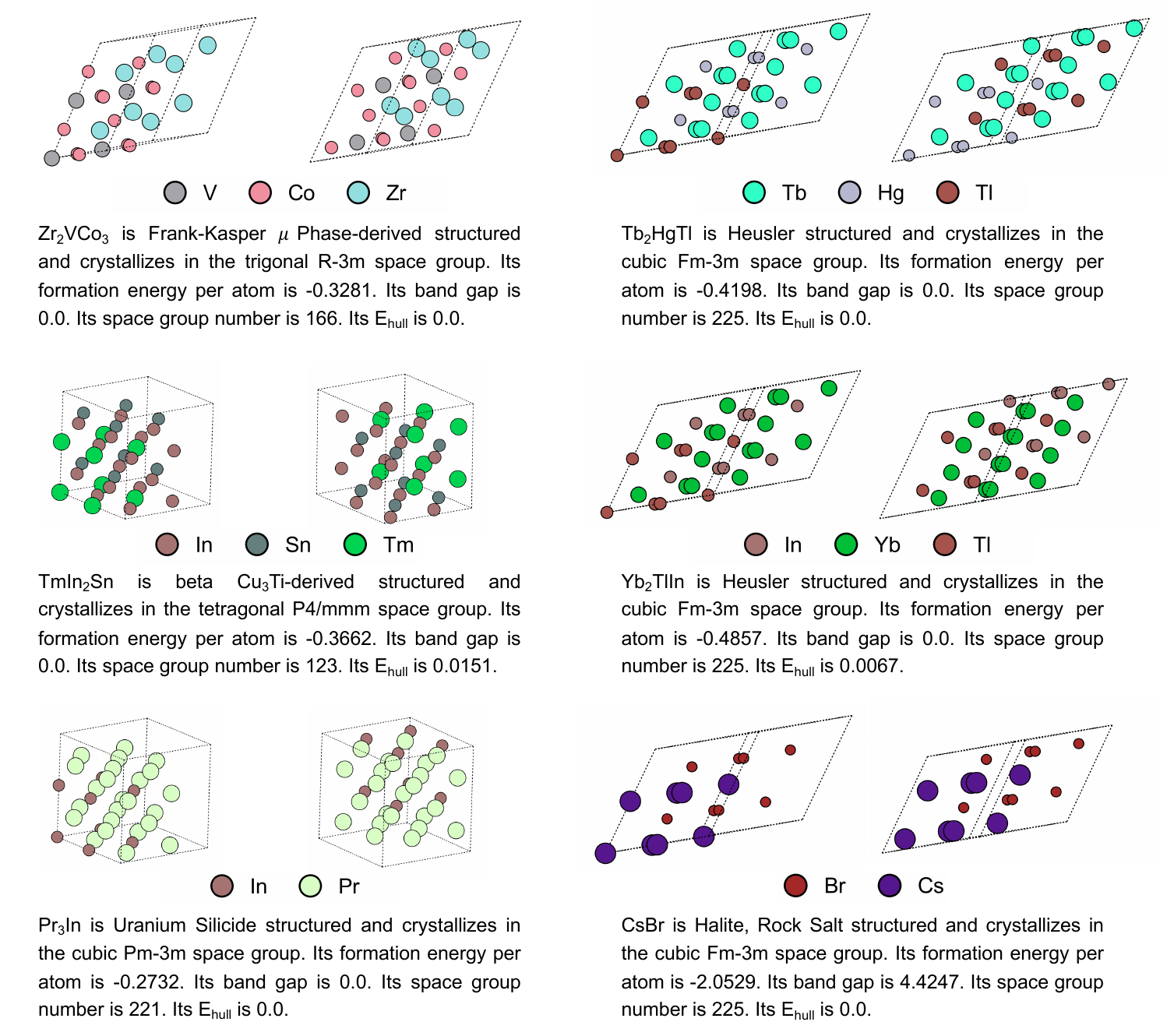}
\caption{\textbf{Representative text-conditioned de novo generations.} Each pair shows a reference MP-20 crystal (left) and a TFMat-generated crystal (right) conditioned on the corresponding structured prompt.}
\label{fig:dng-qualitative}
\end{figure}

Relative to CrystalFlow, TFMat reduces element-count EMD from 0.2489 to 0.1990 and density EMD from 0.1701 to 0.0794, while retaining 99.06\% structural validity and 99.67\% coverage precision (Table~\ref{tab:dng-mp20}). Representative generations are shown in Fig.~\ref{fig:dng-qualitative}. These changes indicate that the language condition biases population-level statistics without degrading geometric plausibility. The remaining bottleneck is atom-type generation, as reflected by weaker compositional validity than the unconditional baseline and the strongest published text-guided diffusion result.

\begin{table}[!htbp]
\centering
\caption{\textbf{MP-20 text-conditioned de novo generation.} Values are percentages except for EMD metrics, where lower is better.}
\label{tab:dng-mp20}
\small
\setlength{\tabcolsep}{4pt}
\begin{tabular}{lcccccc}
\toprule
Model & Comp. valid $\uparrow$ & Struct. valid $\uparrow$ & COV-R $\uparrow$ & COV-P $\uparrow$ & \# Elem. EMD $\downarrow$ & Density EMD $\downarrow$ \\
\midrule
CrystalFlow & \textbf{83.21} & 98.74 & \textbf{99.45} & \textbf{99.76} & 0.2489 & 0.1701 \\
TFMat (Ours) & 81.32 & \textbf{99.06} & 98.91 & 99.67 & \textbf{0.1990} & \textbf{0.0794} \\
\bottomrule
\end{tabular}
\end{table}

\FloatBarrier

\subsection*{Surrogate property diagnostics support composition-selected consistency}
The MP-20 prompts also contain formation-energy and band-gap cues, so we examined whether these coarse signals survive generation when composition is reasonably consistent. This is a surrogate diagnostic rather than a population-wide property benchmark. A composition-selected subset of 2,000 TFMat DNG candidates was evaluated with CGCNN regressors trained on MP-20 formation energy and band gap.\cite{XieGrossman2018} The selection procedure, element-match distribution and outlier rule are reported in Supplementary Note 4.

\begin{figure}[!tbp]
\centering
\includegraphics[width=\textwidth,height=0.50\textheight,keepaspectratio]{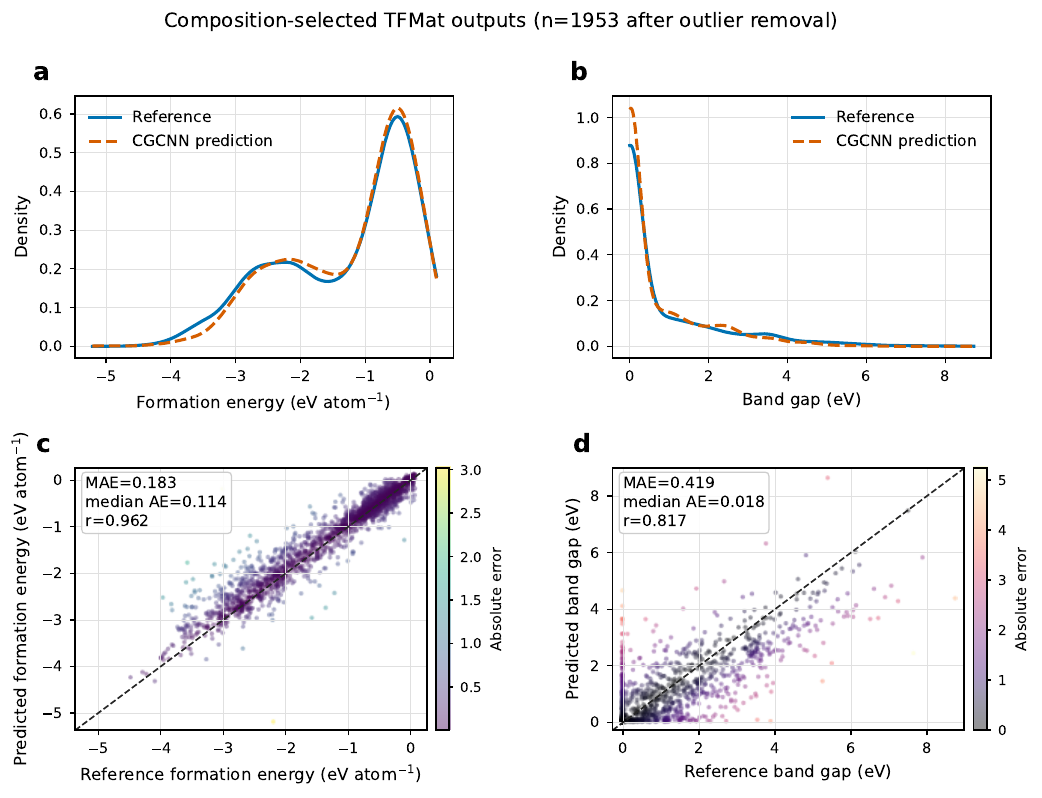}
\caption{\textbf{Surrogate property diagnostics for the composition-selected subset.} Kernel density estimates and scatter panels compare reference prompt values with CGCNN-predicted properties for selected TFMat outputs.}
\label{fig:property-validation}
\end{figure}

Within this selected subset, CGCNN predictions were available for 1,994 structures. Formation-energy sign agreement is 95.94\% (1,913/1,994). After removing 41 numerical outliers with absolute formation-energy error above 10 eV atom$^{-1}$, the cleaned formation-energy MAE is 0.183 eV atom$^{-1}$, the median absolute error is 0.114 eV atom$^{-1}$ and Pearson $r=0.962$ (Fig.~\ref{fig:property-validation}). Band-gap zero/nonzero type agreement is 83.65\%, with MAE 0.419 eV, median absolute error 0.018 eV and Pearson $r=0.817$. These values support prompt-level consistency for the composition-selected subset, especially for formation-energy scale. They do not establish DFT stability, synthesizability or unfiltered population-wide property control.

\subsection*{Graph-embedding diagnostics support the population shift}
\begin{figure}[!tbp]
\centering
\includegraphics[width=0.49\textwidth]{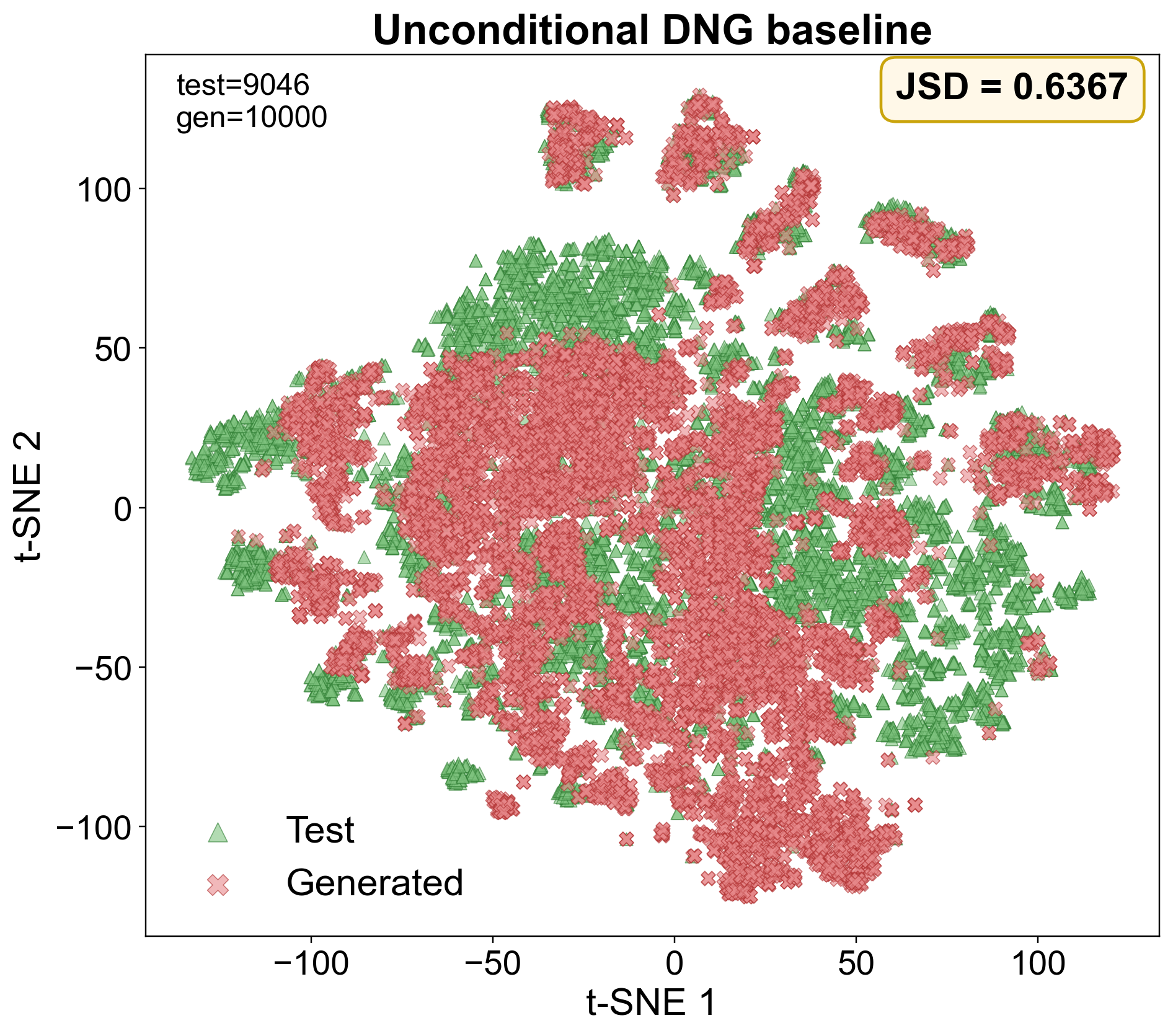}\hfill
\includegraphics[width=0.49\textwidth]{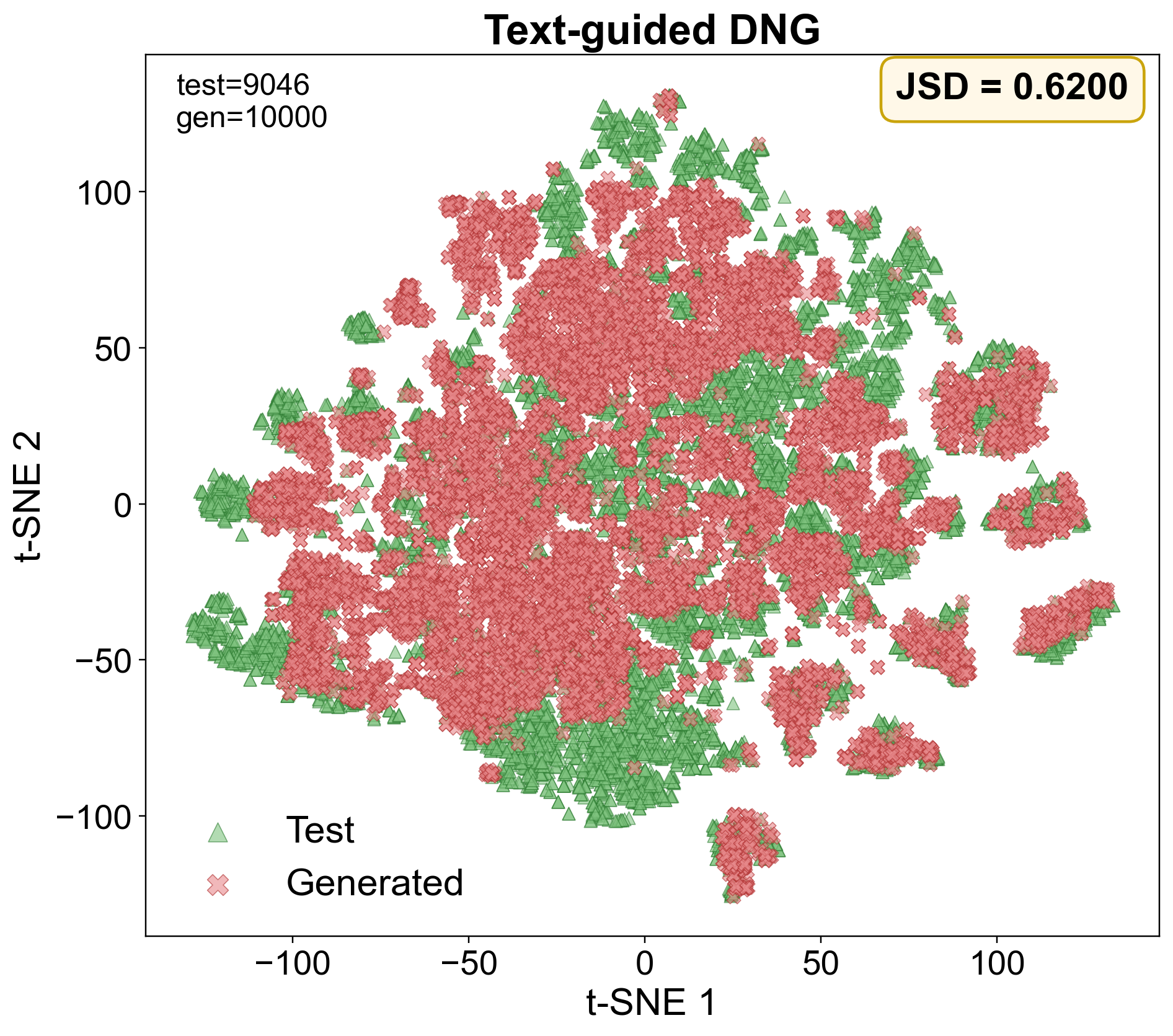}
\caption{\textbf{t-SNE diagnostic for unconditional and text-conditioned MP-20 generation.} Left: unconditional DNG baseline. Right: text-conditioned DNG. Each panel contains 9,046 MP-20 test structures and 10,000 generated structures and uses its own t-SNE projection.}
\label{fig:tsne-comparison}
\end{figure}

The population-level picture is consistent with this interpretation. We embedded MP-20 test crystals and generated crystals with a pretrained graph encoder, then projected each generated set together with the MP-20 test set using t-SNE (Fig.~\ref{fig:tsne-comparison}). The unconditional CrystalFlow baseline and TFMat both cover broad regions of the test distribution, as expected from their shared geometric backbone. The text-conditioned run nevertheless shows a modest improvement in the per-panel histogram diagnostic, with Jensen-Shannon divergence decreasing from 0.6367 to 0.6200. A separate shared-coordinate diagnostic gives the same direction of change, with coverage recall increasing from 0.453 to 0.473, manifold precision increasing from 0.500 to 0.512 and shared t-SNE Jensen-Shannon divergence decreasing from 0.435 to 0.429 (Supplementary Table~\ref{tab:si-tsne-metrics}).

The visualization is not a definitive metric because t-SNE distances are projection-dependent and the two displayed panels are projected separately. Its value is instead as a consistency check on the DNG statistics. TFMat changes the generated population most clearly through marginal distribution alignment, not through a large jump in validity, and the generated points remain spread across multiple clusters. This reduces the likelihood that the improved EMD values arise from a visually obvious collapse to a single prototype.

\FloatBarrier

\section*{Discussion}
TFMat supports a bounded conclusion: structured materials text can steer a flow-matching crystal generator towards more useful regions of crystal space. The strongest evidence is the one-sample CSP result, where the text-conditioned model improves match rate across Perov-5, Carbon-24 and MP-20 while retaining CrystalFlow as the central geometric generator. This is the regime in which semantic priors should be most visible, because the model has only one chance to choose a plausible structural basin.

The comparison with TGDMat clarifies what kind of progress this represents. TGDMat remains a strong text-guided diffusion baseline, particularly for RMSE and MP-20 DNG compositional validity. TFMat is therefore not a universal replacement for diffusion-based text guidance. Its contribution is complementary: a compact language condition can be coupled to a continuous flow-matching generator and can improve sample-efficient structural retrieval. The MP-20 pattern, where match rate improves more clearly than RMSE, suggests that the current model is better at reaching relevant basins than at completing final geometric refinement.

The DNG and diagnostic results extend, but also limit, this claim. Improvements in element-count and density EMDs, a modest shift in graph-embedding overlap, and composition-selected formation-energy and band-gap consistency all point in the same direction: structured text changes the generated distribution in ways that are not captured by validity alone. At the same time, compositional validity trails the best published text-guided diffusion result, and the property analysis depends on CGCNN surrogates and a selected subset. These observations make atom-type generation, exact prompt adherence and unfiltered property consistency the most important technical bottlenecks.

The present prompts also define an important scope boundary. They are structured database-derived descriptions that include formula or composition-level information, space-group descriptors and scalar property values. The results therefore demonstrate controlled metadata-conditioned generation, not independent discovery from unconstrained natural language. The MP-20 prompt-control experiments support a prompt-content effect, but broader field ablations, unseen-prototype tests and stronger out-of-distribution prompt checks are needed to distinguish semantic generalization from reconstruction using highly informative prompts. Density-functional relaxation and unfiltered-population analysis will also be required before making claims about thermodynamic stability of newly generated materials.\cite{Hohenberg1964DFT,Kohn1965DFT}

Within these limits, the result suggests a practical design principle for future crystal generators. Text should not be treated only as annotation attached after generation; when embedded and fused into a geometric flow, structured materials descriptions can become an actionable prior over the search trajectory. This connects the benchmark result to larger closed-loop discovery workflows, where human hypotheses, database records, language-model planners and autonomous platforms need a shared representation of intent. TFMat is one step in that direction: it improves a current crystal-generation task while preserving an interface that can be inspected, edited and reused by both researchers and automated materials-science systems.

\section*{Methods}
\subsection*{Datasets and text conditions}
We evaluated TFMat on Perov-5, Carbon-24 and MP-20, following the benchmark splits used by CDVAE, DiffCSP and TGDMat.\cite{Xie2022CDVAE,Jiao2023DiffCSP,Das2025TGDMat} In CSP, the atom types and composition are fixed and the model predicts lattice and fractional coordinates. In DNG, the model generates atom types, lattice and fractional coordinates jointly after sampling the number of atoms from the empirical training distribution.

Text conditions were generated from structured metadata fields associated with each crystal record. For MP-20, these prompts include formula-level identity, crystal-system or space-group information, formation energy, band gap and energy above hull when available. MatSciBERT embeddings were precomputed with mean pooling before training. The model therefore receives a structured semantic condition, not raw coordinates.

\subsection*{Flow-matching formulation for CSP}
TFMat follows the flow-matching view of generative modeling, in which a neural network learns a velocity field that transports samples from a simple source distribution to the data distribution.\cite{Lipman2023FlowMatching,Liu2023RectifiedFlow,Albergo2023StochasticInterpolants,Tong2024OTCFM} This deterministic transport perspective is complementary to denoising diffusion and score-based generative modeling.\cite{Ho2020DDPM,Song2021ScoreSDE,Karras2022EDM} For a crystal with $N$ atoms and fixed atom types $\mathbf{a}=(a_1,\ldots,a_N)$, the generated state is
\begin{equation}
\mathbf{x} = (\mathbf{k}, \mathbf{F}),
\end{equation}
where $\mathbf{k}\in\mathbb{R}^{6}$ is a six-dimensional lattice-polar representation and $\mathbf{F}\in[0,1)^{N\times3}$ contains fractional coordinates. The target crystal is $\mathbf{x}_1=(\mathbf{k}_1,\mathbf{F}_1)$. We sample a source state
\begin{equation}
\mathbf{k}_0 \sim \mathcal{N}(\mathbf{0}, \sigma_k^2\mathbf{I}), \qquad
\mathbf{F}_0 \sim \mathcal{U}([0,1)^{N\times3}).
\end{equation}
Fractional-coordinate displacement is wrapped with the minimum-image convention,
\begin{equation}
\Delta \mathbf{F} = ((\mathbf{F}_1-\mathbf{F}_0-0.5)\bmod 1)-0.5.
\end{equation}
For $t\sim\mathcal{U}(0,1)$, the interpolation is
\begin{equation}
\mathbf{k}_t = \mathbf{k}_0+t(\mathbf{k}_1-\mathbf{k}_0), \qquad
\mathbf{F}_t = \mathbf{F}_0+t\Delta\mathbf{F}.
\end{equation}
The target velocities are
\begin{equation}
\mathbf{v}^{\star}_{k}=\mathbf{k}_1-\mathbf{k}_0,\qquad
\mathbf{v}^{\star}_{F}=\Delta\mathbf{F}.
\end{equation}
The network predicts $(\hat{\mathbf{v}}_{k},\hat{\mathbf{v}}_{F})$ from $(\mathbf{k}_t,\mathbf{F}_t,\mathbf{a},t,\mathbf{c})$, where $\mathbf{c}$ is the projected text condition. The training loss is
\begin{equation}
\mathcal{L}_{\mathrm{CSP}} =
\lambda_k \|\hat{\mathbf{v}}_{k}-\mathbf{v}^{\star}_{k}\|_2^2
+ \lambda_F \sum_{i=1}^{N}\|\hat{\mathbf{v}}_{F,i}-\mathbf{v}^{\star}_{F,i}\|_2^2.
\end{equation}

\subsection*{Text-conditioned sampling}
Sampling starts from the source state and integrates the learned velocity field with Euler updates. With $T$ integration steps and $\Delta t=1/T$,
\begin{equation}
\mathbf{k}_{t+\Delta t}=\mathbf{k}_{t}+\Delta t\,\hat{\mathbf{v}}_{k}, \qquad
\mathbf{F}_{t+\Delta t}=(\mathbf{F}_{t}+\Delta t\,\hat{\mathbf{v}}_{F})\bmod 1.
\end{equation}
For text-conditioned models, the default conditional sampling factor is 1.0. Thus, the main reported TFMat runs should be interpreted as text-conditioned flow sampling. Additional guidance-factor sweeps vary the interpolation between conditional and unconditional decoder calls, in the broad spirit of classifier-free guidance,\cite{Ho2022ClassifierFreeGuidance} but the present manuscript does not claim an isolated guidance contribution.

\subsection*{Extension to de novo generation}
For DNG, the state includes a relaxed atom-type representation,
\begin{equation}
\mathbf{x}=(\mathbf{Q},\mathbf{k},\mathbf{F}),
\end{equation}
where $\mathbf{Q}\in\mathbb{R}^{N\times d_t}$ is decoded to discrete atom types at the end of sampling. The trajectory is
\begin{equation}
\mathbf{Q}_t=\mathbf{Q}_0+t(\mathbf{Q}_1-\mathbf{Q}_0),\quad
\mathbf{k}_t=\mathbf{k}_0+t(\mathbf{k}_1-\mathbf{k}_0),\quad
\mathbf{F}_t=\mathbf{F}_0+t\Delta\mathbf{F}.
\end{equation}
The model predicts the joint velocity field $(\hat{\mathbf{v}}_{Q},\hat{\mathbf{v}}_{k},\hat{\mathbf{v}}_{F})$, and the DNG training loss is
\begin{equation}
\mathcal{L}_{\mathrm{DNG}} =
\lambda_Q \|\hat{\mathbf{v}}_{Q}-\mathbf{v}^{\star}_{Q}\|_2^2
+ \lambda_k \|\hat{\mathbf{v}}_{k}-\mathbf{v}^{\star}_{k}\|_2^2
+ \lambda_F \sum_{i=1}^{N}\|\hat{\mathbf{v}}_{F,i}-\mathbf{v}^{\star}_{F,i}\|_2^2.
\end{equation}

\subsection*{Evaluation metrics}
CSP match rate and RMSE were computed with the \texttt{pymatgen} \texttt{StructureMatcher}\cite{Ong2013Pymatgen} using a site tolerance of 0.5, an angle tolerance of $10^{\circ}$ and a lattice tolerance of 0.3. In the 20-candidate evaluation, a test case is counted as matched if at least one generated candidate matches the reference. DNG metrics follow the CDVAE/DiffCSP/TGDMat protocol: compositional validity, structural validity, coverage recall (COV-R), coverage precision (COV-P), and Earth mover's distances for property or distribution statistics.\cite{Xie2022CDVAE,Jiao2023DiffCSP,Das2025TGDMat} All percentages in tables are reported on a 0--100 scale.

\subsection*{Visualization and property diagnostics}
The CSP and DNG qualitative panels were assembled from ASE renderings. The t-SNE diagnostics use pretrained graph-encoder embeddings of MP-20 test and generated structures. Each displayed t-SNE panel contains 9,046 test structures and 10,000 generated structures; a separate shared-coordinate t-SNE analysis is reported in the Supplementary Information. The property diagnostic used CGCNN regressors trained on MP-20 formation energy and band gap. The diagnostic subset was selected by ranking generated structures by element-composition match score; it is therefore a targeted consistency analysis rather than an unbiased estimate over all generated crystals.

\section*{Data availability}
The Perov-5, Carbon-24 and MP-20 benchmark datasets are publicly available through the CDVAE benchmark resources.\cite{Xie2022CDVAE} The derived text annotations, precomputed MatSciBERT embeddings, model checkpoints, generated structures and evaluation outputs are available from the TFMat dataset archive at \url{https://huggingface.co/datasets/littlepeachs/TFMat}.

\section*{Code availability}
The TFMat implementation and the scripts required to reproduce the reported generation, evaluation and figures are available at \url{https://github.com/littlepeachs/TFMat}.

\section*{Competing interests}
The authors declare no competing interests.

\FloatBarrier

\FloatBarrier
\clearpage
\appendix
\section*{Supplementary Information}
\setstretch{1.0}
\captionsetup{skip=2pt}
\renewcommand{\thetable}{S\arabic{table}}
\renewcommand{\thefigure}{S\arabic{figure}}
\setcounter{table}{0}
\setcounter{figure}{0}

\subsection*{Supplementary Note 1: Dataset description}
The benchmark tasks use the standard Perov-5, Carbon-24 and MP-20 crystal-generation splits adopted by prior crystal generative models. Perov-5 is a chemically constrained perovskite benchmark, Carbon-24 tests covalent carbon-network generation with larger unit cells, and MP-20 is the broadest benchmark used here, covering inorganic Materials Project structures with up to 20 atoms per unit cell. Table~\ref{tab:si-datasets} reports the split sizes used for evaluation.

The two evaluation regimes use the same data resources but impose different generation constraints. In CSP, atom types and composition are fixed and the model predicts lattice parameters and fractional coordinates. In DNG, the model generates atom types, lattice and fractional coordinates jointly after sampling the number of atoms from the empirical training distribution. The MP-20 t-SNE diagnostic uses 9,046 held-out structures, which is why the visualization count differs from the 8,000-sample CSP evaluation split.

\begin{table}[!htbp]
\centering
\caption{\textbf{Benchmark split sizes and conditioning fields.} The MP-20 t-SNE diagnostic uses 9,046 held-out structures.}
\label{tab:si-datasets}
\small
\setlength{\tabcolsep}{5pt}
\begin{tabularx}{\textwidth}{lrrrX}
\toprule
Dataset & Train & Validation & Test & Text fields used for conditioning \\
\midrule
Perov-5 & 11,356 & 3,787 & 3,785 & Formula, composition, prototype and symmetry-derived descriptors where available \\
Carbon-24 & 6,090 & 2,032 & 2,030 & Formula, carbon framework descriptors and symmetry-derived descriptors where available \\
MP-20 & 27,131 & 9,046 & 8,000 & Formula, crystal system or space group, formation energy, band gap and energy above hull where available \\
\bottomrule
\end{tabularx}
\end{table}

\subsection*{Supplementary Note 2: Model and method details}
TFMat keeps the CrystalFlow flow-matching backbone as the central generator and adds a structured semantic condition derived from database text. This design makes the comparison with CrystalFlow interpretable: the main architectural change is the conditioning pathway rather than a replacement of the geometric transport model. The condition vector is injected into the network as a fixed materials-language representation projected into the model conditioning space.

The text conditions used in this work are generated from structured database fields rather than manually written prose. This choice makes the conditioning channel reproducible and keeps the comparison with CrystalFlow interpretable. Each prompt is assembled from fields that a materials scientist would plausibly use when specifying a target: formula or composition, prototype or structural family when available, crystal system or space-group descriptors, and selected scalar property values. The prompts intentionally do not contain fractional coordinates or lattice vectors.

Each structured prompt is paired with a MatSciBERT representation. Embeddings are computed with a frozen MatSciBERT encoder followed by mean pooling over token representations. The resulting vector is projected into the TFMat conditioning space by a small multilayer perceptron before being fused with the flow-matching backbone. Freezing the text encoder reduces the number of trainable parameters and makes the experiment closer to a controlled test of whether existing materials-language representations can act as useful priors for crystal generation.

Table~\ref{tab:si-metric-definitions} summarizes the metrics used across CSP and DNG. The match-rate and RMSE metrics are reference-based and therefore apply most directly to CSP, where a target structure is known for each composition. The DNG metrics are population-level diagnostics: they measure validity, coverage and distributional similarity, but they do not by themselves establish exact prompt recovery, thermodynamic stability or synthesizability.

\begin{table*}[!htbp]
\centering
\caption{\textbf{Metric definitions and interpretation.} The table separates reference-based CSP metrics from population-level DNG diagnostics.}
\label{tab:si-metric-definitions}
\small
\setlength{\tabcolsep}{5pt}
\begin{tabularx}{\textwidth}{l>{\raggedright\arraybackslash}X>{\raggedright\arraybackslash}X}
\toprule
Metric & What it measures & Main limitation \\
\midrule
CSP match rate & Fraction of test cases for which at least one generated candidate matches the reference under \texttt{StructureMatcher} thresholds & Sensitive to matcher thresholds and to the number of generated candidates \\
CSP RMSE & Geometric error for matched structures after structure matching & Only meaningful for matched structures; it does not penalize unmatched failures directly \\
Compositional validity & Fraction of generated compositions passing charge or composition-validity checks in the benchmark protocol & Does not guarantee exact formula recovery from a text prompt \\
Structural validity & Fraction of generated structures passing basic geometric validity checks & Does not guarantee stability after DFT relaxation \\
COV-R & Coverage recall of the generated distribution relative to the test set & Depends on fingerprint choice and distance threshold \\
COV-P & Coverage precision of generated structures relative to the test distribution & Can remain high even when some prompt-level attributes are not exactly recovered \\
EMD statistics & Earth mover's distance between generated and reference distributions for selected scalar or count features & Summarizes marginal distributions and may miss conditional mismatches at the individual-prompt level \\
t-SNE JSD & Histogram divergence between low-dimensional embedding distributions & Projection-dependent and therefore used only as a qualitative diagnostic \\
\bottomrule
\end{tabularx}
\end{table*}

This distinction explains why the manuscript reports several diagnostics rather than a single score. CSP match rate tests whether text conditioning improves the chance of entering the correct structural basin. DNG validity and EMD metrics test whether the generated population is plausible and distributionally aligned. The composition-selected CGCNN diagnostic asks a narrower question: among candidates that already agree well with the target element set, do generated structures preserve coarse property cues from the prompt? The answer is positive for formation-energy sign and moderate for band-gap type, but this remains a surrogate-model diagnostic.

\subsection*{Supplementary Note 3: CSP experiments and prompt-control analyses}
The main text emphasizes the one-sample regime because it is the most direct test of sample efficiency. Table~\ref{tab:si-csp-comparison} reports both one-sample and 20-sample comparisons between CrystalFlow and TFMat. The gain is largest when the model has only one chance to sample a structure, especially on Perov-5 and Carbon-24. Under 20 evaluations, TFMat remains strongest on MP-20, while Carbon-24 and Perov-5 show a tradeoff in which TFMat gives lower RMSE but a slightly lower match rate than CrystalFlow.

\begin{table*}[!htbp]
\centering
\caption{\textbf{Full CSP results for CrystalFlow and TFMat.} Match rates are percentages. $\Delta$ match is TFMat minus CrystalFlow in percentage points.}
\label{tab:si-csp-comparison}
\small
\setlength{\tabcolsep}{4pt}
\resizebox{\textwidth}{!}{%
\begin{tabular}{llccccc}
\toprule
Dataset & Evaluations & CrystalFlow match & CrystalFlow RMSE & TFMat match & TFMat RMSE & $\Delta$ match \\
\midrule
Perov-5 & 1 & 53.69 & 0.0953 & 91.33 & 0.0449 & +37.65 \\
Carbon-24 & 1 & 15.02 & 0.3095 & 46.40 & 0.1353 & +31.38 \\
MP-20 & 1 & 67.65 & 0.0791 & 77.80 & 0.0868 & +10.15 \\
Perov-5 & 20 & 100.00 & 0.0416 & 96.62 & 0.0349 & -3.38 \\
Carbon-24 & 20 & 89.21 & 0.2208 & 85.02 & 0.1561 & -4.19 \\
MP-20 & 20 & 85.38 & 0.0614 & 92.04 & 0.0665 & +6.66 \\
\bottomrule
\end{tabular}%
}
\end{table*}

The prompt conditions used for CSP are structured rather than free-form. Table~\ref{tab:si-prompt-fields} lists the fields and their intended roles. The structured-prompt design is best viewed as a controlled materials-metadata interface: it is more reproducible than natural prose, but it is weaker than coordinate conditioning because it does not provide fractional coordinates or lattice vectors.

\begin{table*}[!htbp]
\centering
\caption{\textbf{Prompt fields used as structured text conditions.} Field availability varies across datasets and records. The table describes the intended semantic role of each field rather than a guarantee that every record contains every field.}
\label{tab:si-prompt-fields}
\small
\setlength{\tabcolsep}{5pt}
\begin{tabularx}{\textwidth}{l>{\raggedright\arraybackslash}X>{\raggedright\arraybackslash}X}
\toprule
Prompt field & Semantic role & Important limitation \\
\midrule
Formula or composition & Defines the chemical identity or composition-level target and gives the model a strong chemical prior & Highly informative for CSP; shuffled-prompt and field-ablation controls are needed to quantify leakage \\
Prototype or structural family & Provides human-readable structural context when available & Not uniformly available across all records \\
Crystal system or space group & Encodes coarse symmetry information that may constrain lattice and site patterns & Exact symmetry recovery is not guaranteed by population-level DNG metrics \\
Formation energy & Provides a coarse stability cue for property-aware generation & Used as a scalar text value, not as a DFT relaxation target during sampling \\
Band gap & Provides an electronic-type cue, especially zero/nonzero gap information & Band-gap prediction is intrinsically noisier than formation-energy prediction in the CGCNN diagnostic \\
Energy above hull & Provides a metastability-related descriptor when available & Missing values and database uncertainty limit its use as a strict generation target \\
\bottomrule
\end{tabularx}
\end{table*}

\FloatBarrier

Prompt-control experiments on MP-20 test whether the improvement depends on prompt-content alignment rather than only on extra conditioning parameters (Table~\ref{tab:si-csp-prompt-controls}). A milder property-shuffle control keeps formula and symmetry fields fixed while replacing scalar property values from chemically related donor prompts; it retains most of the full-text gain, consistent with formula and symmetry being the dominant CSP cues. A partial embedding perturbation replaces 20\% of prompt embeddings by randomly shuffled embeddings and performs close to the no-text baseline, showing that embedding-level corruption is less interpretable and more disruptive than field-level prompt edits.

\begin{table*}[!htbp]
\centering
\caption{\textbf{MP-20 CSP prompt-control experiments.} All rows use 100 ODE steps, coordinate annealing and one generated candidate per test case. The full-text, property-shuffle and partial-shuffle rows use the text-conditioned CSP model; no-text uses the CrystalFlow baseline.}
\label{tab:si-csp-prompt-controls}
\small
\setlength{\tabcolsep}{5pt}
\resizebox{\textwidth}{!}{%
\begin{tabular}{lccc}
\toprule
Condition & Prompt modification & Match rate $\uparrow$ & RMSE $\downarrow$ \\
\midrule
Full structured prompt & None & 0.771875 & 0.087776 \\
Property-shuffle control & Formula and symmetry retained; scalar property fields shuffled & 0.743500 & 0.088601 \\
Partial-shuffle control & 20\% of prompt embeddings replaced by shuffled embeddings & 0.655750 & 0.092532 \\
No-text baseline & No text condition & 0.669750 & 0.077648 \\
\bottomrule
\end{tabular}%
}
\end{table*}

\FloatBarrier

\subsection*{Supplementary Note 4: De novo generation experiments and hyperparameter selection}
The best TFMat DNG result does not arise from simply increasing the number of ODE steps. We compared four targeted sampling configurations, varying integration depth, coordinate-annealing slope and guidance factor while keeping the evaluation pipeline fixed (Table~\ref{tab:si-sampling-sweep}). The best balance is obtained by the 100-step configuration with coordinate-annealing slope 10 and the default text-conditioned sampling factor. This setting gives the best number-of-elements EMD and density EMD while retaining high structural validity and coverage precision.

\begin{table*}[!htbp]
\centering
\caption{\textbf{Targeted TFMat sampling sweep on MP-20 DNG.} Best values are bold and second-best values are underlined. The default conditional sampling factor is 1.0.}
\label{tab:si-sampling-sweep}
\small
\setlength{\tabcolsep}{4pt}
\resizebox{\textwidth}{!}{%
\begin{tabular}{ccccccccc}
\toprule
ODE steps & Anneal slope & Guide factor & Comp. validity $\uparrow$ & Struct. validity $\uparrow$ & COV-R $\uparrow$ & COV-P $\uparrow$ & \# Element EMD $\downarrow$ & Density EMD $\rho \downarrow$ \\
\midrule
100 & 5 & 0.8 & 80.34 & \textbf{99.35} & \underline{99.50} & \underline{99.51} & 0.3178 & \underline{0.1397} \\
100 & 5 & 1.2 & 81.03 & 97.55 & 99.43 & 99.42 & 0.2428 & 0.2102 \\
100 & 10 & 1.0 & \textbf{81.32} & \underline{99.06} & 98.91 & \textbf{99.67} & \textbf{0.1990} & \textbf{0.0794} \\
200 & 5 & 1.0 & \underline{81.11} & 98.30 & \textbf{99.55} & 99.27 & \underline{0.2208} & 0.2659 \\
\bottomrule
\end{tabular}%
}
\end{table*}

The sweep suggests that the text-conditioned flow benefits from a moderate integration depth paired with a sharper coordinate schedule. Increasing the number of ODE steps to 200 slightly improves coverage recall but worsens density alignment. Fixed guidance factors above or below the default also fail to dominate. For this reason, the main MP-20 DNG result uses the 100-step, anneal-slope-10 configuration.

The displayed t-SNE panels in Fig.~\ref{fig:tsne-comparison} are per-panel projections: each generated set is embedded together with the same MP-20 test set and then projected independently. This makes the plots suitable for visual inspection of overlap within each panel, but it means that point coordinates should not be compared directly between the two panels. To provide a more quantitative cross-model check, Table~\ref{tab:si-tsne-metrics} reports both the per-panel Jensen-Shannon divergence shown in the figure and a separate shared-coordinate t-SNE diagnostic.

\begin{table}[!htbp]
\centering
\caption{\textbf{MP-20 graph-embedding t-SNE diagnostics.} The per-panel JSD values correspond to the two panels in Fig.~\ref{fig:tsne-comparison}. The shared-coordinate metrics are computed from a single projection containing the test set and both generated sets.}
\label{tab:si-tsne-metrics}
\small
\setlength{\tabcolsep}{5pt}
\resizebox{\textwidth}{!}{%
\begin{tabular}{lcccccc}
\toprule
Model & Test count & Generated count & Per-panel JSD $\downarrow$ & Shared COV-R $\uparrow$ & Shared COV-P $\uparrow$ & Shared JSD $\downarrow$ \\
\midrule
Unconditional DNG baseline & 9,046 & 10,000 & 0.636697 & 0.453350 & 0.4997 & 0.435103 \\
Text-guided DNG & 9,046 & 10,000 & 0.619971 & 0.472806 & 0.5124 & 0.428576 \\
\bottomrule
\end{tabular}%
}
\end{table}

The property diagnostic in the main text uses a selected subset rather than the full generated population. The 2,000 structures were selected from the first 5,000 TFMat DNG candidates by element-composition match score, so the result measures the most composition-consistent subset of the generated population. Within this selected subset, 1,350 structures perfectly match the target element set and no structure has zero element overlap (Table~\ref{tab:si-element-match}). The average element-match score is 0.9075.

\begin{table}[!htbp]
\centering
\caption{\textbf{Element-composition match distribution for the top 2,000 generated structures.} Structures are ranked by element-composition match score before the CGCNN property diagnostic is performed.}
\label{tab:si-element-match}
\small
\setlength{\tabcolsep}{8pt}
\begin{tabular}{lccc}
\toprule
Match category & Count & Percentage & Score range \\
\midrule
Complete match & 1,350 & 67.50\% & [1.0] \\
Partial match & 650 & 32.50\% & (0.0, 1.0) \\
No match & 0 & 0.00\% & [0.0] \\
\midrule
\textbf{Overall average} & \textbf{2,000} & \textbf{100.00\%} & \textbf{0.9075} \\
\bottomrule
\end{tabular}
\end{table}

CGCNN predictions were available for 1,994 selected structures. The surrogate validation errors are summarized in Table~\ref{tab:si-cgcnn-training}. We removed 41 numerical formation-energy outliers with absolute error above 10 eV atom$^{-1}$ before computing the cleaned formation-energy scatter statistics reported in Fig.~\ref{fig:property-validation}. No outlier removal was applied to the band-gap diagnostic. These choices should be treated as diagnostic filtering rather than as a benchmark protocol.

\begin{table}[!htbp]
\centering
\caption{\textbf{CGCNN surrogate models used in the property diagnostic.} Validation errors document the reliability of the surrogate models rather than replacing first-principles validation.}
\label{tab:si-cgcnn-training}
\small
\setlength{\tabcolsep}{6pt}
\begin{tabular}{lccc}
\toprule
Target property & Best validation MAE & Best epoch & Diagnostic use \\
\midrule
Formation energy & 0.0391 eV atom$^{-1}$ & 84/100 & Sign agreement and cleaned regression statistics \\
Band gap & 0.3747 eV & 6/100 & Zero/nonzero agreement and regression statistics \\
\bottomrule
\end{tabular}
\end{table}

The large difference between formation-energy and band-gap validation errors is consistent with the main-text diagnostic: formation energy shows stronger correlation and lower median error, whereas band gap is more reliable as a coarse electronic-type indicator than as a precise regression target. This is why the main text emphasizes formation-energy sign and band-gap zero/nonzero agreement rather than claiming quantitative electronic-structure accuracy.

A separate diagnostic run with 1,000 generated samples, text embeddings, 100 ODE steps, coordinate-annealing slope 5 and a conditional sampling factor of 1.0 is not used as the main DNG result because its formula, space-group and crystal-system matching rates are very low: 2.10\%, 2.30\% and 0.00\%, respectively. It is retained here because it defines an important interpretation boundary: strong population-level validity and coverage metrics do not guarantee exact recovery of the prompt formula or symmetry labels in a small diagnostic sample. This motivates the composition-selected property analysis above and the cautionary language in the Discussion.

\FloatBarrier

\end{document}